# Time dependent spatial entanglement in atom-field interaction

Ivan P. Christov

Physics Department, Sofia University, 1164 Sofia, Bulgaria

Email: ivan.christov@phys.uni-sofia.bg

**Abstract**

By using stochastic ensembles of walkers in physical and in one-body Hilbert spaces the recently proposed time-dependent quantum Monte Carlo (TDQMC) method offers the unique capability to calculate one-body density matrices at fully correlated level, without referencing the many-body quantum state. Here TDQMC is applied to study entanglement of simple systems such as Moshinsky atom (oscillator potentials) and atoms with Coulomb potentials in one spatial dimension. Our findings indicate that the dynamic entanglement of atoms exposed to powerful ultrashort laser pulse can be easily manipulated by introducing an appropriate phase modulation where the negative chirp enhances the entanglement while the positive one suppresses it. These findings can be used to explore the correlation properties of different constituents of complex quantum systems subjected to appropriately shaped laser radiation.

1. Introduction

Entanglement is a fundamental concept of atomic physics which plays an important role in quantum mechanics of systems involving more than one particle. In such systems entanglement implies inseparability of the various degrees of freedom which is also responsible for the phenomenon of quantum non-locality. Besides its important role in quantum information which is based essentially on manipulation of entangled states [1], entanglement can also be considered as a basic ingredient for description of the structural properties of composite quantum systems [2], for e.g. quantifying the correlations between the different species in the system. Although quantum entanglement is most often considered in spin systems where it can be treated by relatively simple means, there is a growing interest to the position space (spatial) entanglement which occurs in systems with more than one degree of freedom. In particular, the correlations induced by different interactions may lead to spatial entanglement which manifests as an inseparability of the many-body wave function and it is expected to be important for exploring basic properties of the ground states of atoms and molecules and also their evolution in space and time. Previous investigations of quantum entanglement in atomic systems include two-electron model atoms, helium atoms and ions, and molecules [3]–[7]. The use of precise Kinoshita-type wave functions for the ground state and some excited states of helium [4] and using other configuration interaction variational methods [8] has allowed to determine the linear and von Neumann entanglement entropies for these relatively simple systems. The study of spatial entanglement may also help to clarify the intimate mechanisms behind the quantum-to-classical transition [9].

In order to address the computational difficulties related to the exponential scaling of the quantum many-body problems various approximation have been employed to treat the quantum entanglement in atomic systems which include density functional theory (DFT) [10] and density matrix renormalization group (DMRG) [11,12]. These methods, however, consider in most cases the ground state properties of the system under consideration. Besides the entanglement of ground state and low-energy eigenstates, entanglement evolution has been investigated in low-dimensional scattering [13] and for dissociating diatomic molecules [14]. The rapid progress in laser technologies in recent years has provided sources which are capable of delivering highly coherent optical pulses with few-femtosecond duration in visible and sub-femtosecond duration in x-ray domains [15,16]. It is therefore of significant interest to investigate the possibilities to control the dynamics of quantum entanglement in atoms and molecules subjected to such ultra-short electromagnetic pulses where non-perturbative superposition of the ground state and highly excited states can be readily produced and manipulated by different strong-field coherent control techniques [17,18]. In the present paper we focus on the time evolution of quantum entanglement of strongly correlated systems such as one-dimensional atoms exposed to strong ultra-short optical pulses. We use a fully correlated *ab-initio* treatment based on the recently proposed time-dependent quantum Monte Carlo (TDQMC) method which replaces the many-body wave function by a set of entangled one-body wave functions propagating in an effective potential which involves explicitly the spatial quantum non-locality [19]. The method employs particle-wave dichotomy based on generic Hamiltonians instead of mapping the interacting many-body system onto a non-interacting one as done in DFT, and it is easily upgraded to higher spatial dimensions (unlike in DMRG), while scaling almost linearly with the number of particles [20,21]. The paper is structured as follows: the theoretical model to study many-body quantum systems and their interactions is presented in Sec. 2 and in the Appendix; in Sec. 3 we first benchmark the TDQMC predictions for atoms of up to ten spinless electrons against the well-known Moshinsky atom and for atoms with long-range (Coulomb) interactions where exact analytical and/or numerical solutions are available, and then discuss the results for the time evolution of the linear entropy as an entanglement measure for these atoms in an external field; finally in Sec. 4 we present the conclusions.

## 2. Theoretical model

### 2.1 Preliminaries

Here we employ the time dependent quantum Monte Carlo as a stochastic method to circumvent the exponentially large Hilbert space of the many-body quantum problem by introducing particle-wave dichotomy through statistical ensemble of Monte Carlo (MC) walkers in physical space and a concurrent ensemble of walkers in one-body Hilbert space (guide waves), for each quantum particle (for details see the Appendix, and [19-21]). Then, the probability density for i-th physical particle in space is represented by a cloud of M MC walkers with coordinates $\mathbf{r}_i^k(t)$ where i=1,2,…,N, and k=1,2,…,M. The corresponding guide waves $\varphi_i^k(\mathbf{r}_i,t)$ obey one-body Schrodinger-type of equations in an external potential:

$$i\hbar \frac{\partial}{\partial t}\varphi_i^k(\mathbf{r}_i,t) = \left[-\frac{\hbar^2}{2m_i}\nabla_i^2 + V_{eff}^k(\mathbf{r}_i,t) + V_{ext}(\mathbf{r}_i,t)\right]\varphi_i^k(\mathbf{r}_i,t), \tag{1}$$

where $V_{eff}^k(\mathbf{r}_i,t)$ is the effective potential experienced by $\varphi_i^k(\mathbf{r}_i,t)$ due to the MC walkers which belong to the rest of the particles in the system, which is given by a MC convolution:

$$V_{eff}^k(\mathbf{r}_i,t) = \sum_{j\neq i}^N \frac{1}{Z_j^k} \sum_l^M V\left[\mathbf{r}_i,\mathbf{r}_j^l(t)\right] K\left[\mathbf{r}_j^l(t),\mathbf{r}_j^k(t),\sigma_j^k\right], \tag{2}$$

where (see Eqs.(A.7)-(A.8)):

$$K\left[\mathbf{r}_j,\mathbf{r}_j^k(t),\sigma_j^k\right] = \exp\left(-\frac{\left|\mathbf{r}_j - \mathbf{r}_j^k(t)\right|^2}{2\sigma_j^k\left(\mathbf{r}_j^k,t\right)^2}\right) \tag{3}$$

is a kernel function which determines the degree of spatial non-locality quantified by the characteristic non-local length $\sigma_j^k(\mathbf{r}_j^k,t)$. The connection between the trajectories $\mathbf{r}_i^k(t)$ and the corresponding guide waves $\varphi_i^k(\mathbf{r}_i,t)$ is given by the de Broglie-Bohm guiding equations for real-time propagation:

$$\mathbf{v}_i^k(t) = \frac{\hbar}{m_i}\text{Im}\left[\frac{\nabla_i \varphi_i^k(\mathbf{r}_i,t)}{\varphi_i^k(\mathbf{r}_i,t)}\right]_{\mathbf{r}_i = \mathbf{r}_i^k(t)}, \tag{4}$$

and it is given by a drift-diffusion process for the ground-state preparation (imaginary-time propagation), Eqs.(A.19)-(A.20):

$$d\mathbf{r}_i^k(\tau) = \mathbf{v}^{Dk}_i d\tau + \mathbf{\eta}_i(\tau)\sqrt{\frac{\hbar}{m_i}d\tau}, \tag{5}$$

where:

$$\mathbf{v}^{Dk}_i(\tau) = \frac{\hbar}{m_i}\frac{\nabla_i \varphi_i^k(\mathbf{r}_i,\tau)}{\varphi_i^k(\mathbf{r}_i,\tau)} \tag{6}$$

is the drift velocity and $\mathbf{\eta}(\tau)$ is Markovian stochastic process whose amplitude tends to zero toward steady state [21]. In fact, it is seen from Eq.(6) and Eq.(4) that the drift velocities for imaginary time and for real time propagation are described by similar expressions except that for the former the drift velocity depends on the gradient of the real field while for the latter it depends on the gradient of the phase of a complex field.

Thus, the set of TDQMC equations (1)-(6) can be considered as a method to self-consistently evolve stochastic ensembles of particles and guide waves in space and time. Since thousands of replicas of the guide waves are propagated under random effective potentials in Eq.(1), the fully correlated picture of many-body quantum dynamics can be recovered in TDQMC. During the ground state preparation stage, the imaginary time propagation of the TDQMC equations, Eqs.(1)-(6), can be considered as an iterative procedure where the trajectories of the walkers for a given electron are substituted back into the effective potential $V_{eff}^k(\mathbf{r}_i,t)$ at the RHS of the Eq.(1) at each time step (successive improvements of the potential) until consistent distributions of particles and guide waves are established towards steady state. These iterations are reminiscent of the original Hartree self-consistent field method (e.g. [22]) applied here to solve for mixed states which involve both wave functions and particles. Then, for a good performance of the algorithm one should make sure that the smoothed distribution of MC walkers in space for the i-*th* particle is as close as possible to $\sum_{k=1}^{M}\left|\varphi_i^k(\mathbf{r}_i,t)\right|^2 / M$ which, as we shall see below, is nothing but the diagonal of the one-body density matrix in spatial representation.

Clearly, in the limiting case where the spatial non-local length $\sigma_j^k \to 0$, the TDQMC equations (1)-(4) are transformed into a set of local Schrödinger equations, which neglects the trajectory entanglement (ultra-correlated dynamics, [21]), while for $\sigma_j^k \to \infty$ the Hartree approximation is restored which neglects the wave function entanglement. It can be expected therefore that there exist an optimal value of $\sigma_j^k$ which ensures correct entanglement of both kinds and at the same time it minimizes the energy of the system. First, we clarify the physical meaning of $\sigma_j^k$ in Eq.(3) by assuming as a main approximation that it is proportional to a global quantity such as the standard deviation $\sigma_j(t)$ of the Monte Carlo sample which characterizes the spatial uncertainty of the walker distribution for a given quantum species:

$$\sigma_j^k\left(\mathbf{r}_j^k,t\right)=\alpha_j.\sigma_j(t), \tag{7}$$

In TDQMC $\alpha_j$ is considered to be a variational parameter which is determined by minimizing the ground state energy of the system and which may depend inversely on the average distance between the walkers for the different particles [20]. Note that for two and three dimensions, $\sigma_j(t)$ should be related to the variance-covariance matrix of the Monte Carlo sample. It is important to stress also that since the TDQMC equations are optimizing between the ultra-correlated [21] and the un-correlated (Hartree) cases, which are available analytically and describe unitary evolution, the real-time TDQMC dynamics is also expected to be unitary, which is essential for building correct density matrices as shown below.

### 2.2 Density matrices and entanglement measures

As we have seen there are two concurrent statistical ensembles to describe each physical particle in TDQMC: an ensemble of walkers in physical space and a concurrent ensemble of guide waves in the one-body Hilbert space where in the mean-field approximation all walkers are guided by the same guide wave and that particle is therefore in a pure state. In the general case, however, each guide wave for a given particle experiences different effective interaction potential $V_{eff}^k(\mathbf{r}_i,t)$ in Eq.(1) which results in a mixture of guide waves for that particle. That mixture can be used next to efficiently build one-body density matrix considered as variance-covariance matrix for the random variables $\varphi_i^k(\mathbf{r},t)$ in the one-body Hilbert space. Most importantly, this can be done directly from the TDQMC output without the need to calculate the density matrix of the whole system and next reduce it to one-body density matrix by e.g. multi-dimensional integration that would be much more computationally demanding. In our case, the time dependent one-body density matrix for the *i*-th particle can be easily calculated from [24]:

$$\rho_i(\mathbf{r}_i,\mathbf{r}_i',t) = \frac{1}{M}\sum_{k=1}^{M}\varphi_i^{k*}(\mathbf{r}_i,t)\varphi_i^k(\mathbf{r}_i',t) \quad, \tag{8}$$

Once calculated, the one-body density matrix of Eq.(8) can be further used to quantify coherence and entanglement properties for parts of a complex quantum system without referencing to the many-body quantum state, as we shall see in Part 3 below.

In order to calculate the spatial entanglement for the ground state and for real time dynamics here we use the linear entropy obtained from the one-body density matrix, Eq.(8), which in fact quantifies the mixedness of the reduced density matrix:

$$S_L^i(t) = 1 - Tr(\rho_i^2) = 1 - \int \rho_i^2(\mathbf{r}_i,\mathbf{r}_i,t)d\mathbf{r}_i \tag{9}$$

Another measure for the entanglement and correlation is provided by inverse purity function [25]:

$$S_L^{i\prime}(t) = \frac{1}{Tr(\rho_i^2)} = \frac{1}{\int \rho_i^2(\mathbf{r}_i,\mathbf{r}_i,t)d\mathbf{r}_i} \tag{10}$$

### 2.3 Quantum - semi-classical - classical transition

It is clear from Eqs.(1)-(6) that under certain conditions the set of TDQMC trajectories may exhibit semi-classical and even classical behavior. For example, whenever the distributions $\left|\varphi_i^k(\mathbf{r}_i,t)\right|^2$ remain strongly localized around some mean trajectory:

$$\bar{\mathbf{r}}_i(t) = \int |\varphi_i(\mathbf{r}_i,t)|^2 \mathbf{r}_i d\mathbf{r}_i = \frac{1}{M}\sum_{k}^{M} \mathbf{r}_i^k(t) \tag{11}$$

during the whole time evolution of the system, we arrive at the semi-classical limit of TDQMC where all guide waves experience the same mean field potential due to that trajectory [26]:

$$V_{eff}^k(\mathbf{r}_i,t) = \sum_{j\neq i}^{N} V\left[\mathbf{r}_i,\bar{\mathbf{r}}_j(t)\right], \tag{12}$$

which corresponds to uncorrelated guide waves by setting $\sigma_j^k \to \infty$ in Eqs.(1)-(3). On the other hand, by allowing that $\hbar/m_i \to 0$ in Eq.(1) the quantum diffusion in Eq. (5) vanishes and so does the non-local length $\sigma_j^k(\mathbf{r}_j^k,t)$ together with the width of the distribution $|\varphi_i^k(\mathbf{r}_i,t)|^2$. As a result, the ensemble of walkers for each particle collapses to a single walker thus transforming the set of quantum particles to a set of classical particles with the only force between them being due to the classical potential $V(\mathbf{r}_i,\mathbf{r}_j)$:

$$m_i \frac{d^2\mathbf{r}_i}{dt^2} = -\nabla_i \sum_{j\neq i}^{N} V\left[\mathbf{r}_i,\mathbf{r}_j,t\right]. \tag{13}$$

Classical approaches have been used previously in the context of atomic ionization in strong laser fields where a set of Newton's equations for micro-canonical ensembles of classical Monte Carlo walkers for each electron replace the exact quantum dynamics [27]. Needless to say that the classical calculations cannot provide information needed to explore entanglement and coherence in quantum systems. Other approaches used in microelectronics [28] involve non-Hermitian Hamiltonians which, however, violates the unitarity for real-time dynamics and leads to degradation of both wave functions and density matrices. Also, in order to provide exact quantum trajectories those methods require preliminary guessing of the true many-body wave function, thus facing the exponential scaling of the many-body Schrodinger equation.

## 3. Results and discussion

First, we benchmark the TDQMC predictions against the results from known models which allow exact analytical and/or exact numerical solutions.

### 3.1 Moshinsky atom

The Moshinsky atom consists of N spinless particles harmonically attracted to a nucleus and interacting with each other through a harmonic potential. It is an example of an exactly soluble model which has been widely used to benchmark different approximate many-body approaches such as the quality of Hartree-Fock approximation [29], for exploring low-order density matrices of atomic systems [30] and for investigating entanglement [31,32]. For unit strength of the nuclear attraction the Hamiltonian reads (in one dimension, atomic units):

$$H = \sum_i^N \left( -\frac{1}{2}\frac{\partial^2}{\partial x_i^2} + \frac{1}{2}x_i^2 \right) \pm \frac{\kappa}{2}\sum_{i<j}^N (x_i - x_j)^2 , \tag{14}$$

where the parameter $\kappa$ determines the strength of mutual interaction and the minus/plus sign corresponds to mutual repulsion/attraction, respectively. The exact ground state density matrix is obtained to be [33]:

$$\rho_e(x_1, x_2) = \left[ \frac{\delta_N N / \pi}{N - 1 + \delta_N} \right]^{1/2} \exp\left[ -a_1(x_1^2 + x_2^2) + a_2 x_1 x_2 \right], \tag{15}$$

where:

$$\delta_N = \sqrt{1 + \kappa N} , \tag{16}$$

$$a_1 = \frac{1}{4N} \frac{\left[ (N-1)(1+\delta_N^2) + 2(N^2 - N + 1)\delta_N \right]}{N - 1 + \delta_N} , \tag{17}$$

$$a_2 = \frac{1}{2N} \frac{\left[ (N-1)(1-\delta_N)^2 \right]}{N - 1 + \delta_N} , \tag{18}$$

and the exact linear entropy can be calculated from:

$$S_{L,e} = 1 - \int \rho_e^2(x,x) dx, \tag{19}$$

which is to be compared with the numerically determined linear entropy:

$$S_L^i(t) = 1 - \int \rho_i^2(x_i,x_i,t) dx_i; \quad i=1,....N, \tag{20}$$

where:

$$\rho_i^2(x_i,x_i',t) = \int \rho_i(x_i,x,t)\rho_i(x,x_i',t) dx, \tag{21}$$

and, from Eq.(8) we have:

$$\rho_i(x_i,x_i',t) = \frac{1}{M}\sum_{k=1}^{M} \varphi_i^{k*}(x_i,t)\varphi_i^k(x_i',t), \tag{22}$$

where $\varphi_i^k(x_i,t)$ is obtained from the TDQMC equations:

$$i\frac{\partial}{\partial t}\varphi_i^k(x_i,t) = \left[-\frac{1}{2}\frac{\partial^2}{\partial x_i^2} + \frac{1}{2}x_i^2 + V_{eff}^k(x_i,t)\right]\varphi_i^k(x_i,t), \tag{23}$$

where the effective interaction potential is given by:

$$V_{eff}^k(x_i,t) = \pm\frac{\kappa}{2}\sum_{j\neq i}^{N}\frac{1}{Z_j^k}\sum_{l=1}^{M}\left[x_i - x_j^l(t)\right]^2 \exp\left(-\frac{\left|x_j^l(t) - x_j^k(t)\right|^2}{2\sigma_j^k(x_j^k,t)^2}\right), \tag{24}$$

where:

$$Z_j^k = \sum_{l=1}^{M}\exp\left(-\frac{\left|x_j^l(t) - x_j^k(t)\right|^2}{2\sigma_j^k(x_j^k,t)^2}\right). \tag{25}$$

We start with the calculation of the TDQMC ground state of the atom through minimizing the total energy $E = \sum_{k}^{M} E_L^k$ by varying the parameter $\alpha$ which substitutes for all parameters $\alpha_j$ of Eq.(7) in case of equal spinless particles. This variational optimization results in a ground state energy accurate within three to four significant digits with respect to the exact analytical value of $0.5\left[(N-1)\sqrt{1+\kappa N}+1\right]$ [33]. Figure 1 (a)-(d) shows with blue lines the linear entropy of the ground state of the Moshinsky atom as function of the parameter $\alpha$ for N=1, 3, and 10 particles, respectively, while the red lines represent the exact analytical values for the same number of particles, for $\kappa$=0.2 (attractive potential). The cross-points of the red and the blue lines correspond to the values of $\alpha$ where the true ground state is found, which is verified by the insets showing the optimizing ground state energy versus $\alpha$. Figure 1(d) shows the dependence of the spatial non-local length $\sigma_j^k$ and the optimal value of $\alpha$ on the number of particles. It is seen that both curves decrease monotonically with increasing the number of particles while at the same time the linear entanglement entropy (Fig.1(e)) and the ground state energy (Fig.1(f)) increase. This behavior can be attributed to the attractive force between the particles where the TDQMC results match very well the exact predictions.

Once the ground state is established we next use the guide waves of $\varphi_i^k(x_i,t)$ to numerically calculate the one-body density matrix of Eq.(22), and then follow its real-time evolution. Figure 2 shows with black lines the eigenvalue spectra of the one-body density matrix for the ground state of a Moshinsky atom with N=2, 3, and 10 particles, for the same value of the interaction strength $\kappa$=0.2, to be compared with the exact spectra, from Eq.(15), (red lines). It is seen that the calculated and the exact spectra (occupation numbers) are close down to 16 orders of magnitude and these match almost perfectly down to 6 orders of magnitude. We found that for N=10 particles the leading occupation number for the ground state equals 0.993 for both the exact and TDQMC calculations. Thus, the results from Fig.1 and Fig.2 indicate that despite the large number of guide waves (typically ~10 000) the TDQMC one-body density matrices reproduce remarkably well the exact results.

Since the Moshinsky atom has been treated analytically from time-independent perspective only, we test the real time evolution provided by TDQMC by comparing it to the exact numerical solution of the time-dependent Schrodinger equation, for up to four harmonically interacting particles in a parabolic potential. We calculate the linear entropy for free diffraction (scattering) after the nuclear potential is suddenly switched off after the ground state is established at t=0. Figure 3 shows the linear entropy as function of time for $\kappa$=-0.2 (repulsive potential) for N=2, 3, and 4, where it is seen that the TDQMC results (blue lines) match well the exact ones (red lines) for the whole time evolution.

**3.2 Atoms in external field**

In order to further compare the TDQMC results with the numerically exact solution of the many-body Schrödinger equation with long-range potentials here we use one-dimensional model of N-electron atom [34]:

$$i\frac{\partial}{\partial t}\Psi(x_1,x_2,...,x_N,t) = \left[ -\frac{1}{2}\sum_i^N \frac{\partial^2}{\partial x_i^2} - \sum_i^N \frac{1}{\sqrt{1+x_i^2}} + \sum_{i>j}^N \frac{1}{\sqrt{1+(x_i-x_j)^2}} \right.$$

$$\left. +V_{ext}(x_1,x_2,...,x_N,t) \right]\Psi(x_1,x_2,...,x_N,t). \tag{26}$$

Then, the TDQMC equations read:

$$i\frac{\partial}{\partial t}\varphi_i^k(x_i,t) = \left[ -\frac{1}{2}\frac{\partial^2}{\partial x_i^2} - \frac{N}{\sqrt{1+x_i^2}} + V_{eff}^k(x_i,t) + V_{ext}(x_i,t) \right]\varphi_i^k(x_i,t), \tag{27}$$

$i=1,2,…,N;$ $k=1,…,M,$ and the effective electron-electron potential is given by:

$$V_{eff}^k(x_i,t) = \sum_{j \neq i}^N \frac{1}{Z_j^k} \sum_{l=1}^M \frac{1}{\sqrt{1+\left[x_i - x_j^l(t)\right]^2}} \exp\left( -\frac{\left|x_j^l(t) - x_j^k(t)\right|^2}{2\sigma_j^k\left(x_j^k,t\right)^2} \right), \tag{28}$$

where:

$$Z_j^k = \sum_{l=1}^M \exp\left( -\frac{\left|x_j^l(t) - x_j^k(t)\right|^2}{2\sigma_j^k\left(x_j^k,t\right)^2} \right). \tag{29}$$

The calculation of the ground state of atoms with different N proceeds as before by minimizing the total energy $E = \sum_k^M E_L^k$ through varying the parameter $\alpha$. Since there is no analytical solution in this case, for comparison with the numerically exact results we use the standard definition of the reduced density matrix:

$$\rho(x,x',t) = \frac{\int \Psi(x,x_2,...,x_N,t)\Psi^*(x',x_2,...,x_N,t)dx_2...dx_N}{\int \Psi(x_1,x_2,...,x_N,t)\Psi^*(x_1,x_2,...,x_N,t)dx_1...dx_N} \qquad (30)$$

Because of the exponential scaling of Eq.(26) with the number of particles, here we were able to find numerically exact ground state for up to four spinless electrons. Figure 4(a)-(c) shows with blue lines the linear entropy as a function of $\alpha$ where, similarly to in Section 3.1, these curves cross the exact entropy (red lines) in points where the ground state energies have minimum. Unlike for attractive interaction, here $\alpha_{opt}$ increases with the particle number, while $\sigma_j^k$ decreases, as seen from Fig.4(d). The TDQMC predictions for the linear entropy and for the energy in Fig.4(e)-(f) are again very close to the exact results. Figure 5 presents the most essential results of this work, namely the possibility to control the entanglement entropy by using appropriately modulated electromagnetic pulses. To achieve this we expose the one-dimensional atom to a few cycle laser pulse with carrier frequency 0.092 a.u. and peak intensity 0.12 a.u. ($5.10^{14}$ W/cm$^2$) where the external potential in dipole approximation is given by $V_{ext}(x_i,t) = -x_i E_0(t)\cos\left[\omega_0 t + \gamma t^2\right]$ with $\gamma$ being the rate of the frequency chirp. For $\gamma = 0$ (transform limited pulse) the blue and the red curves in Fig.5(a)-(c) show an increase of the linear entropy with time as the laser pulse shakes the electrons forth and back with respect to the core. Since the square of the density matrix in Eq.(8) mixes all products of guide waves which involve their complex phases acquired during the atom-field interaction, the good correspondence between the TDQMC and the exact results shown in Fig.5 indicate that the quantum coherences are calculated with a very good accuracy by the TDQMC algorithm. By introducing phase modulation with rate $\gamma = \pm 0.00003$ a.u. we observe significant change in the degree of entanglement as shown with green lines in Fig.5 which means that using appropriately shaped laser pulses it might be possible to manipulate the correlations between the electrons. In this calculation, as in the other calculations reported in the present work, the spatial grid spans typically 50 a.u. and a total of ~10 000 walkers and guide waves take part in the TDQMC calculation.

## 4. Conclusions

In this work we have studied the spatial entanglement of interacting particles with oscillator potentials (Moshinsky atom) and with smoothed Coulomb potentials using time-dependent quantum Monte Carlo method. First we benchmark the TDQMC results for the ground state against the exact analytical results available for Moshinsky atom. It was found that the linear entropy as a measure for quantum entanglement in space is correctly predicted by TDQMC for up to 10 particles in Moshinsky atom and the eigenvalue spectra of the corresponding density matrices match very well over several orders of magnitude. The real-time dynamics of entanglement of interacting particles released from the core potential and next experiencing free diffraction show a close correspondence with the numerically exact results obtained from the direct numerical solution of the many-body Schrodinger equation in one spatial dimension. These findings are confirmed next for the ground states of atoms with smoothed Coulomb potentials. The main achievement of the present study is the prediction that the spatial entanglement in atoms can be governed efficiently by applying short electromagnetic pulses with appropriate phase modulation as the negative chirp enhances the entanglement entropy while the positive one suppresses it. The latter can be explained by the excitation of complex superposition of the ground state and excited states where the atomic coherences are determined by the phases of these states. Such a strong-field coherent control may be used to manipulate the correlations between the different species in complex quantum systems such as molecules and clusters for e.g. creation and annihilation of quantum bonds. Clearly, our results are equally applicable to other many-body systems such as Bose-Einstein condensates.

In TDQMC we address the computational problems due to the exponentially increasing Hilbert space of quantum mechanics by using stochastic ensembles of MC walkers in physical and in one-body Hilbert spaces without referencing directly the many-body quantum states. As a result, each walker is guided by its own guide wave and we have not invoked explicitly the symmetry of the guide waves with respect to exchange of the walker's coordinates for indistinguishable particles. At the same time our approach allows us to accurately calculate one body density matrices and pair distributions, which are in fact the main practically important quantities. Unlike in other methods TDQMC uses unitary propagation of the guide waves where the key physical parameter is the non-local quantum correlation length which quantifies the degree of spatial non-locality and is closely related to the quantum uncertainty thus opening new perspectives for studying the relation between the entanglement and uncertainty.

## 5. Acknowledgment

Financial support from the Bulgarian National Science Fund under grant DN18/11 is gratefully acknowledged.

# Appendix

In this appendix we consider the motion of the guide waves, the MC walkers, and the general TDQMC algorithm. For a non-relativistic system of N quantum particles we start with the many-body Schrödinger equation:

$$i\hbar \frac{\partial}{\partial t}\Psi(\mathbf{R},t) = H\Psi(\mathbf{R},t), \tag{A.1}$$

where:

$$H = -\sum_{i}^{N} \frac{\hbar^2}{2m_i}\nabla_i^2 + V(\mathbf{R},t) + V_{ext}(\mathbf{R},t), \tag{A.2}$$

is the many-body quantum Hamiltonian, and $\mathbf{R} = \{\mathbf{r}_1,...,\mathbf{r}_N\}$ are the degrees of freedom. In the typical case the potential $V(\mathbf{R},t)$ in Eq. (A.2) may include electron-nuclear, electron-electron, and nuclear-nuclear potentials, $V_{ext}(\mathbf{R},t)$ is the external potential, and $m_i$ is the mass of $i$-th particle. In general, the Hamiltonian in Eq. (A.2) is not separable in coordinates which is the source for the exponential scaling of Eq. (A.1) with the number of particles. The most popular approach to deal with that scaling is to factorize of the wave function:

$$\Psi(\mathbf{R},t) = \prod_{i=1}^{N} \varphi_i(\mathbf{r}_i,t), \tag{A.3}$$

which reduces Eqs.(A.1),(A.2) to a set of mean field equations for the separate wave functions $\varphi_i(\mathbf{r}_i,t)$, where the motion of each particle occurs in an averaged potential due to the rest of the particles:

$$i\hbar \frac{\partial}{\partial t}\varphi_i(\mathbf{r}_i,t) = \left[ -\frac{\hbar^2}{2m_i}\nabla_i^2 + V^{Hartree}(\mathbf{r}_i,t) + V_{ext}(\mathbf{r}_i,t) \right]\varphi_i(\mathbf{r}_i,t), \tag{A.4}$$

where:

$$V^{Hartree}(\mathbf{r}_i,t) = \sum_{j\neq i}^{N} \frac{1}{\int d\mathbf{r}_j |\varphi_j(\mathbf{r}_j,t)|^2} \int d\mathbf{r}_j |\varphi_j(\mathbf{r}_j,t)|^2 V(\mathbf{r}_i - \mathbf{r}_j), \tag{A.5}$$

with $V(\mathbf{r}_i - \mathbf{r}_j)$ being the pairwise interaction potential for any degree under interest (e.g. for electron-electron Coulomb repulsion). It is clear that the factorization of the many-body wave function in Eq.(A.3) basically ignores the correlation between the different degrees of freedom in Eq.(A.1). The time dependent quantum Monte Carlo (TDQMC) methodology offers a simple and efficient way to recover that correlation by introducing an additional degree of freedom such that the i-th quantum particle is described by a set of M Monte Carlo (MC) walkers in physical space with trajectories $\mathbf{r}_i^k(t)$ (k=1,…M) and a concurrent ensemble of walkers in one-body Hilbert space represented by the wave functions $\varphi_i^k(\mathbf{r}_i,t)$, where the wave functions $\varphi_i^k(\mathbf{r}_i,t)$ play role of guide waves for the walkers. This approach, which we call particle-wave dichotomy in TDQMC, is capable of revealing the correlations (otherwise hidden by the mean-field) by creating a mixture of states for each interacting physical particle. This can be accomplished by e.g. applying a stochastic windowing to the mean-field distribution for the j-the particle $|\varphi_j^k(\mathbf{r}_j,t)|^2$ with a window (kernel) function $K\left[\mathbf{r}_j, \mathbf{r}_j^k(t), \sigma_j^k\right]$ with a width $\sigma_j^k$ centered at the trajectory $\mathbf{r}_j^k(t)$; $j \neq i$, which, for Gaussian kernel, looks like [19-21]:

$$K\left[\mathbf{r}_j, \mathbf{r}_j^k(t), \sigma_j^k\right] = \exp\left(-\frac{|\mathbf{r}_j - \mathbf{r}_j^k(t)|^2}{2\sigma_j^k(\mathbf{r}_j^k,t)^2}\right). \tag{A.6}$$

This introduces a stochastic effective potential which is different for each separate walker pair, and which is given by a Monte Carlo convolution:

$$V_{eff}^k(\mathbf{r}_i,t) = \sum_{j\neq i}^{N} \frac{1}{Z_j^k} \sum_{l}^{M} V\left[\mathbf{r}_i, \mathbf{r}_j^l(t)\right] K\left[\mathbf{r}_j^l(t), \mathbf{r}_j^k(t), \sigma_j^k\right], \tag{A.7}$$

where:

$$Z_j^k = \sum_l^M K\left[\mathbf{r}_j^l(t), \mathbf{r}_j^k(t), \sigma_j^k\right], \tag{A.8}$$

and the characteristic non-local length $\sigma_j^k$ quatifies the degree of spatial non-locality in our model.

This approach transforms the mean-field equation (A.4) into a set of equations for the different replicas $\varphi_i^k(\mathbf{r}_i,t)$ of the one-body wave function $\varphi_i(\mathbf{r}_i,t)$, thus transforming the mean field approximation to a fully correlated picture [19,20]:

$$i\hbar \frac{\partial}{\partial t}\varphi_i^k(\mathbf{r}_i,t) = \left[-\frac{\hbar^2}{2m_i}\nabla_i^2 + V_{eff}^k(\mathbf{r}_i,t) + V_{ext}(\mathbf{r}_i,t)\right]\varphi_i^k(\mathbf{r}_i,t) \tag{A.9}$$

It is important to point out that the connection between $\varphi_i^k(\mathbf{r}_i,t)$ and $\mathbf{r}_i^k(t)$ is twofold in that the particle $\mathbf{r}_i^k(t)$ samples its own distribution given by $|\varphi_i^k(\mathbf{r}_i,t)|^2$ during the ground state preparation (imaginary time propagation, see below), and it is guided by $\varphi_i^k(\mathbf{r}_i,t)$ for real time dynamics trough the de Broglie-Bohm guiding equations:

$$\mathbf{v}_i^k(t) = \frac{\hbar}{m_i}\text{Im}\left[\frac{\nabla_i\varphi_i^k(\mathbf{r}_i,t)}{\varphi_i^k(\mathbf{r}_i,t)}\right]_{\mathbf{r}_i=\mathbf{r}_i^k(t)}, \tag{A.10}$$

where the standard polar decomposition $\varphi_i^k(\mathbf{r}_i,t) = R_i^k(\mathbf{r}_i,t)\exp[iS_i^k(\mathbf{r}_i,t)/\hbar]$ has been substituted in Eq.(A.9), which also leads to a second order equations of motion for the trajectories of the individual walkers which belong to the $i$-th particle [21]:

$$m_i \frac{d^2\mathbf{r}_i^k}{dt^2} = \left\{-\nabla_i\left[-\frac{\hbar^2}{2m_i}\frac{\nabla_i^2 R_i^k(\mathbf{r}_i,t)}{R_i^k(\mathbf{r}_i,t)} + V_{eff}^k(\mathbf{r}_i,t) + V_{ext}(\mathbf{r}_i,t)\right]\right\}_{\mathbf{r}_i^k=\mathbf{r}_i^k(t)}, \tag{A.11}$$

where the de Broglie relation $\mathbf{v}_i^k(t) = \nabla_i S_i^k(\mathbf{r}_i,t)/m_i$ has been used. In order to further reveal the physical meaning of Eq.(A.9) let us compare it with the equation for the exact trajectories stemming from the many-body Schrödinger equation, Eq.(A.1), following the same arguments:

$$m_i \frac{d^2 \mathbf{r}_i^k}{dt^2} = \left\{ -\nabla_i \left[ -\frac{\hbar^2}{2} \sum_{j=1}^N \frac{1}{m_j} \frac{\nabla_j^2 R^k(\mathbf{r}_1,...,\mathbf{r}_j,...,\mathbf{r}_N,t)}{R^k(\mathbf{r}_1,...,\mathbf{r}_j,...,\mathbf{r}_N,t)} + V(\mathbf{r}_1,...,\mathbf{r}_j,...,\mathbf{r}_N,t) + V_{ext}(\mathbf{r}_i,t) \right] \right\}_{\mathbf{r}_j = \mathbf{r}_j^k(t)}. \quad (A.12)$$

Then, it is clear from Eq.(A.11) and Eq.(A.12) that the reduction of the many-body Schrödinger equation to a set of coupled one-body equations, as done in TDQMC, occurs at the price of "shifting" the spatial non-locality from the first term in the RHS of Eq.(A.12), also known as the "quantum potential", to the stochastic effective potential $V_{eff}^k$ in the RHS of Eq.(A.11). In fact, the effective interaction potential which is responsible for the dynamic correlations seems to be a natural place where a significant part of the entanglement should be accommodated. Also, it is seen from Eq.(A.11) that the TDQMC trajectories differ from the exact Bohmian trajectories of Eq.(A.12) which experience the exponential scaling of the many-body Schrödinger equation. At the same time the TDQMC trajectories are determined by first order equations (A.10), thus avoiding the use of quantum potentials.

In order to elucidate further the particle-wave dichotomy in TDQMC here we consider the ground state preparation of a quantum system by searching for a random walk to produce the distribution:

$$f(\mathbf{R},\tau) = |\Psi(\mathbf{R},\tau)|^2, \quad (A.13)$$

where $\Psi(\mathbf{R},\tau)$ obeys the imaginary time many-body Schrödinger equation:

$$-\hbar \frac{\partial}{\partial \tau} \Psi(\mathbf{R},\tau) = \left[ -\frac{\hbar^2}{2m} \nabla^2 + V(\mathbf{R}) \right] \Psi(\mathbf{R},\tau), \quad (A.14)$$

where $\nabla = (\nabla_1, \nabla_2, ..., \nabla_N)$. It can be easily shown that the distribution $f(\mathbf{R}, \tau)$ satisfies a Fokker-Planck-type of equation:

$$\hbar \frac{\partial}{\partial \tau} f(\mathbf{R}, \tau) = \frac{\hbar^2}{2m} \nabla^2 f(\mathbf{R}, \tau) - \hbar \nabla \bullet [\mathbf{v}^D(\mathbf{R}, \tau) f(\mathbf{R}, \tau)] - 2E_L(\mathbf{R}) f(\mathbf{R}, \tau), \qquad (A.15)$$

where:

$$\mathbf{v}^D(\mathbf{R}, \tau) = \frac{\hbar}{m} \frac{\nabla \Psi(\mathbf{R}, \tau)}{\Psi(\mathbf{R}, \tau)} \qquad (A.16)$$

is the drift velocity, and:

$$E_L(\mathbf{R}) = \frac{H\Psi(\mathbf{R}, \tau)}{\Psi(\mathbf{R}, \tau)} = V(\mathbf{R}) - \frac{\hbar^2}{2m} \frac{\nabla^2 \Psi(\mathbf{R}, \tau)}{\Psi(\mathbf{R}, \tau)} \qquad (A.17)$$

is the local energy. Hence, the Langevin equation which corresponds to Eq.(A15) reads [23]:

$$d\mathbf{R}(\tau) = \mathbf{v}^D(\mathbf{R}(\tau)) d\tau + \boldsymbol{\eta}(\tau) \sqrt{\frac{\hbar}{m} d\tau}, \qquad (A.18)$$

where $\boldsymbol{\eta}(\tau)$ is a vector random variable with zero mean and unit variance. After the factorization of Eq.(A.3) we obtain from Eq.(A.16)-(A.18) the working equations for the TDQMC trajectories, for imaginary-time propagation:

$$\mathbf{v}^{Dk}_i(\tau) = \frac{\hbar}{m_i} \frac{\nabla_i \varphi_i^k(\mathbf{r}_i, \tau)}{\varphi_i^k(\mathbf{r}_i, \tau)}, \qquad (A.19)$$

for the drift velocity, which determines the drift-diffusion process:

$$d\mathbf{r}_i^k(\tau) = \mathbf{v}_i^{Dk} d\tau + \mathbf{\eta}_i(\tau)\sqrt{\frac{\hbar}{m_i}d\tau}, \tag{A.20}$$

and for the local energy:

$$E_L^k = \sum_{i=1}^{N}\left[-\frac{\hbar^2}{2m_i}\frac{\nabla_i^2 \varphi_i^k(\mathbf{r}_i^k,\tau)}{\varphi_i^k(\mathbf{r}_i^k,\tau)} + \sum_{j>i}^{N} V(\mathbf{r}_i^k,\mathbf{r}_j^k)\right]_{\substack{\mathbf{r}_i^k = \mathbf{r}_i^k(\tau) \\ \mathbf{r}_j^k = \mathbf{r}_j^k(\tau)}}, \tag{A.21}$$

where the k-th walker $\mathbf{r}_i^k(\tau)$ is biased to yield the distribution $|\varphi_i^k(\mathbf{r}_i,\tau)|^2$ instead of $|\Psi(\mathbf{R},\tau)|^2$.

The TDQMC algorithm starts with initial ensembles of M random walkers and M guide waves for each physical particle. Then, at each imaginary time step one calculates the effective potentials, Eqs.(A.7), (A.8) and then moves the guide waves $\varphi_i^k(\mathbf{r}_i,\tau)$ by solving the one-body Schrodinger equations (A.9), while at the same time moves the concurrent walkers $\mathbf{r}_i^k(\tau)$ by Metropolis sampling of $|\varphi_i^k(\mathbf{r}_i,\tau)|^2$ plus (eventually) drift, Eq.(A.9), which leads to new effective potentials, etc. The effective potentials are efficiently calculated from the trajectories $\mathbf{r}_i^k(\tau)$ where due to the randomness of these trajectories their number in Eq.(A.7) can be much less than the total number of walkers M, which significantly speeds up the calculation. Similarly to diffusion Monte Carlo (DMC) multiplication of walkers (branching) can be employed to improve the performance of the algorithm where a weighting is used to control birth and death of walkers and guide waves according to their excess local energy $E_L^k - E_T$, $E_T$ being a trial energy. However, the TDQMC algorithm was found to give good results also in case of no branching where the drift term in Eq.(A.18) should be neglected, which could serve as a good starting point for the ground state variational optimization. An additional benefit of using TDQMC is that no multi-parametric Slater-Jastrow wave functions are involved, unlike in DMC or in variational Monte Carlo calculations [23].

**Figure captions**

**FIG. 1.** (a)-(c) Linear entropies as function of $\alpha$ for N=2, 4, and 10 particles in Moshinsky atom; blue lines: TDQMC calculation, red lines: exact result. The insets show the ground-state energy as function of $\alpha$. (d) spatial non-local length (red line) and optimal $\alpha_{opt}$ (blue line) as function of particle number; (e)-(f) linear entropy and energy as function of particle number. Blue lines: TDQMC results, red lines: exact results.

**FIG. 2.** Eigenvalue spectrum $\lambda_k(k)$ of the one-body density matrix for N=2, 3, and 10 particles in Moshinsky atom. Black lines: TDQMC result, red lines: exact result.

**FIG. 3.** Linear entropy for N=2, 3, and 4 particles released from Moshinsky atom as function of time during diffraction. Blue lines: TDQMC result, red lines: numerically exact result.

**FIG. 4.** (a)-(c) Linear entropies as function of $\alpha$ for N=2, 3, and 4 spinless electrons in an atom; blue lines: TDQMC calculation, red lines: exact numerical result; (d) spatial non-local length (red line) and optimal $\alpha_{opt}$ (blue line) as function of particle number; (e)–(f) linear entropy and atomic energy as function of particle number. Blue lines: TDQMC result, red lines: exact numerical result.

**FIG. 5.** Linear entropy as function of time for atom exposed to short optical pulse with no phase modulation. Blue lines: TDQMC results, red lines: exact numerical result. The green lines show the results for pulse with phase modulation where the signs next to the curves denote the sign of the frequency modulation, for N=2, 3, and 8 electrons.

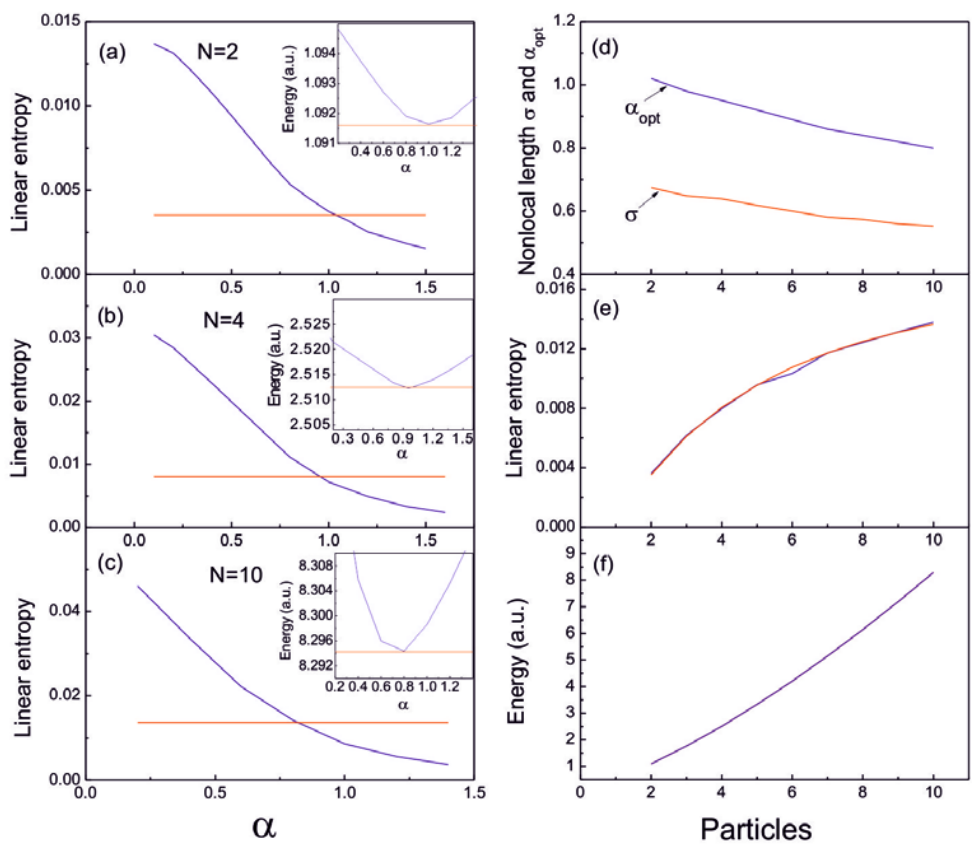

**FIG. 1**

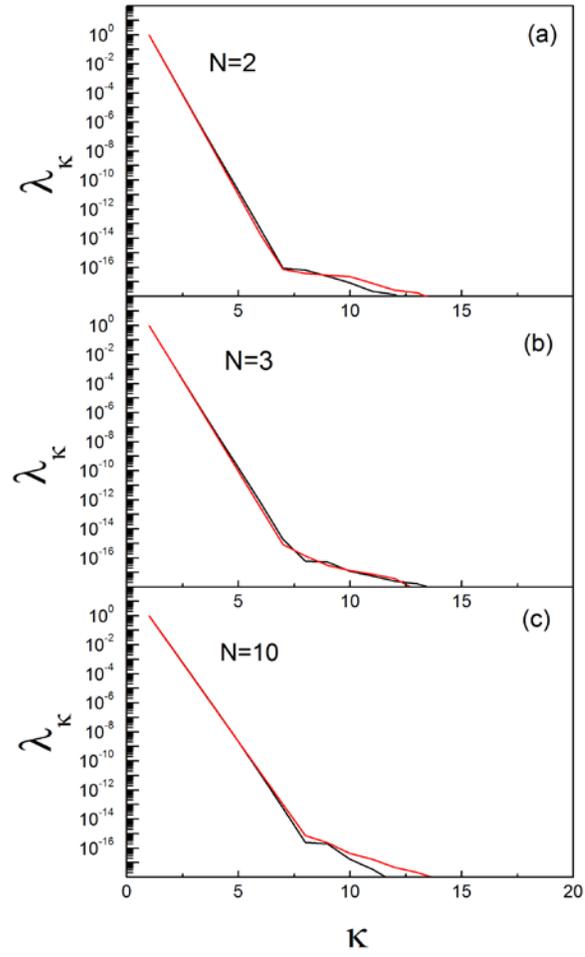

**FIG. 2**

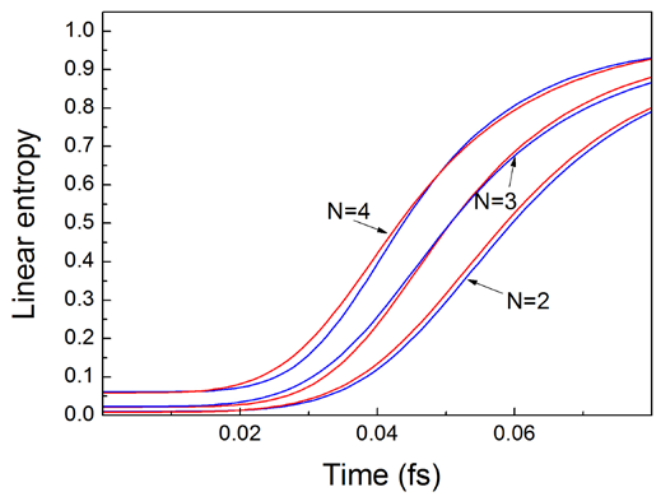

**FIG. 3**

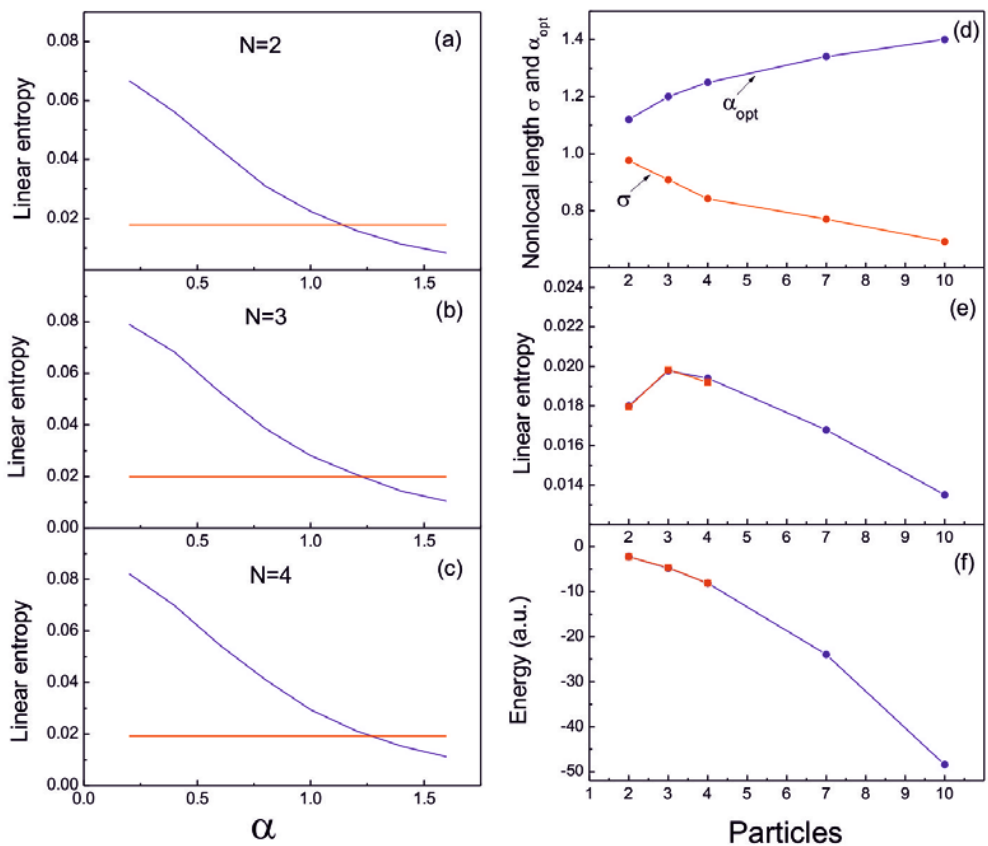

**FIG. 4**

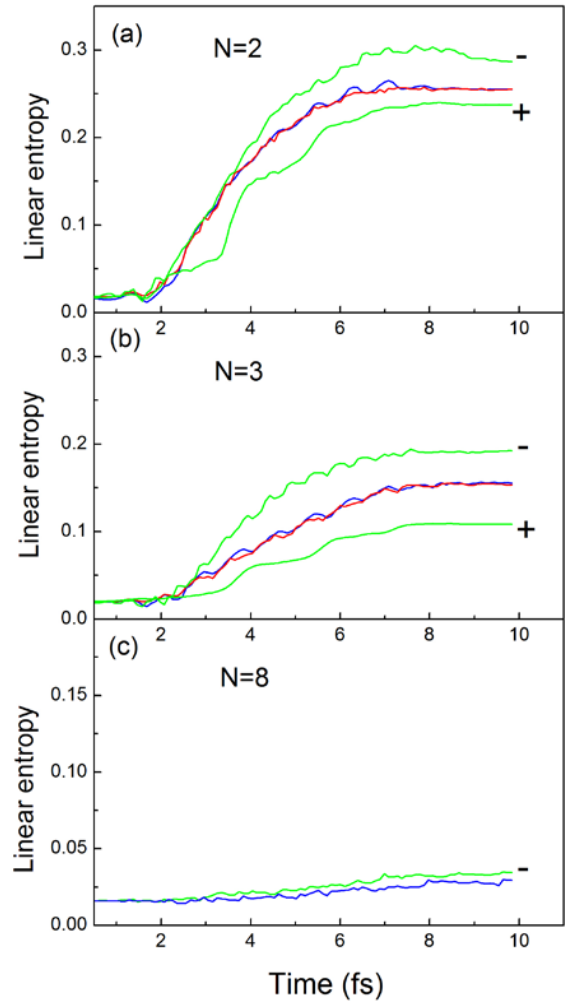

**FIG. 5**